\documentclass[12pt]{iopart}
\usepackage{amssymb}
\usepackage{graphicx}
\usepackage{epstopdf}
\usepackage{epsfig}
\usepackage{color}
\begin{document}

\title{Traveling Solitons in Long-Range Oscillator Chains}

\author{George Miloshevich$^{1}$ Jean Pierre Nguenang$^{2}$ Thierry Dauxois$^{3}$ Ramaz Khomeriki$^{4}$
and Stefano Ruffo$^{5}$}

\address{${\ }^1$Department of Physics, The University of
Texas
at Austin, Austin TX 78712, USA \\
${\ }^2$Fundamental Physics Laboratory: Group of Nonlinear Physics
and Complex Systems, Department of Physics, University of Douala,
P.O. Box 24157, Douala, Cameroon \\
${\ }^3$Univ. Lyon, ENS Lyon, Univ. Claude Bernard, CNRS,
Laboratoire
de Physique, F-69342 Lyon, France\\
${\ }^4$Department of Physics, Faculty of Exact and Natural
Sciences, Tbilisi State University, 0128 Tbilisi,
Georgia\\
${\ }^5$SISSA and INFN, via Bonomea 265, 34136 Trieste, Italy}

\begin{abstract}

\end{abstract}We investigate the existence and propagation of solitons in a
long-range extension of the quartic Fermi-Pasta-Ulam (FPU) chain of
anharmonic oscillators. The coupling in the linear term decays as a
power-law with an exponent $1 < \alpha \leq 3$. We obtain an
analytic perturbative expression of traveling envelope solitons by
introducing a Non Linear Schr\"odinger (NLS) equation for the slowly
varying amplitude of short wavelength modes. Due to the non analytic
properties of the dispersion relation, it is crucial to develop the
theory using discrete difference operators. Those properties are
also the ultimate reason why kink-solitons may exist but are
unstable, at variance with the short-range FPU model. We
successfully compare these approximate analytic results with
numerical simulations.


\section{Introduction}
The study of the equipartition process in the Fermi-Pasta-Ulam-Tsingou
(FPU) model of nonlinearly coupled oscillators
~\cite{FPU,galavotti,tod} has led to important discoveries in both
statistical mechanics \cite{izrailev, ruffo1} and nonlinear science
\cite{scott}. At the same time, nonlinear oscillator chains serve as
the simplest prototypes for complex condensed matter
systems \cite{flach1,flach2} and biophysical phenomena
\cite{takeno,thierry2}. In particular, the study of FPU chains
has historically motivated the discovery of solitons~\cite{zabusky,thierry1}.
Further developments, namely nonintegrable (Klein-Gordon~\cite{klein} and Frenkel-Kontorova~\cite{braun})
and integrable Toda~\cite{toda} chains helped much in understanding
the interplay between integrability and chaos~\cite{FPU}. These concepts have been applied
to describe transport properties in electric transmission lines~\cite{trans} and even in
quantum systems, such as Josephson junction parallel arrays and lattices \cite{zolot,ramaz3}.
In most cases, the analysis was restricted to one-dimensional
($d=1$) lattices where oscillators interact only with nearest
neighbors, i.e. to short-range interactions.

In recent years there has been a growing interest in systems with long-range
interactions~\cite{long,Campabook}. In such systems, either the
two-body potential or the coupling at separation $r$ decays with a
power-law $r^{-\alpha}$. When the power $\alpha$ is less than the
dimension of the embedding space $d$, these systems violate
additivity, a basic feature of thermodynamics, leading to unusual
properties like ensemble inequivalence, broken ergodicity,
quasistationary states.

Long-range coupled oscillator models have been previously introduced
to cope with dipolar interactions in mechanistic DNA
models~\cite{dna}. They describe also ferroelectric~\cite{electr}
and magnetic~\cite{magnet} systems, where the long-range coupling is
provided again by dipolar forces. Other candidates for application
are cold gases: dipolar bosons~\cite{trom,atomdipol}, Rydberg
atoms~\cite{rydberg}, atomic ions~\cite{kastner,ions}. Moreover, one
can mention optical wave turbulence~\cite{optics} and scale-free
avalanche dynamics \cite{ava}, where such long-range couplings
appear.
The extension of the FPU problem to include long-range couplings is
rarely considered~\cite{flach,frac,helena} and attention has been
mainly focused on deriving the continuous counterpart of the discrete long
range models~\cite{frac}, on considering thermalization
properties caused by the long-range character of the interactions
\cite{helena,milo}, or on finding conditions for the existence of
standing localized solutions like breathers~\cite{flach}.

In this Letter, we consider a generalization of the FPU model by
introducing a long-range coupling in the linear term decaying with
the power $\alpha$, while keeping the nonlinear term short-range. We
have chosen the power in the range $1<\alpha\leq 3$, where we obtain
qualitatively similar results. Below $\alpha=1$ the energy diverges
and above $\alpha=3$ the systems becomes short-range. Dipolar
systems correspond to the power $\alpha=3$, while the power
$\alpha=2$ has been considered for crack front propagation along
disordered weak planes between solid blocks \cite{ava} and contact
lines of liquid spreading on solid surfaces~\cite{liquid}.

\section{Methods}
The Hamiltonian of the model reads
\begin{equation}
{\cal H} = \sum\limits_{j=-N}^N \frac{p_j^2}{2} +
\frac{1}{2}\sum_{j>\ell=1}^N \frac{(u_j-u_\ell)^2}{{| j-\ell
|}^\alpha}+\sum\limits_{\ell=1}^N
\frac{\left(u_{n+1}-u_n\right)^4}{4}
\end{equation}
and the corresponding equations of motion are
\begin{equation} \label{first}
\ddot{u}_\ell = \sum_{{j=1 \\ j \ne \ell}}^N \frac{u_j-u_\ell}{{|
j-\ell
|}^\alpha}+\left(u_{\ell+1}-u_\ell\right)^3+\left(u_{\ell+1}-u_\ell\right)^3
\end{equation}

Assuming plane wave solutions of the form
\begin{equation}
u_\ell =  (A/2)e^{i(q\ell-\Omega t)}+c.c.\label{plane}
\end{equation}
and substituting them into the equations of motion, we obtain the nonlinear
dispersion relation in Rotating Wave Approximation (RWA) for the
normal mode frequencies $\Omega_n$ and wave numbers $q_n$
\begin{equation}
\left(\Omega_n\right)^2 = 2\sum_{m=1}^N \frac{1-\cos(q_n
m)}{m^\alpha}+3|A|^2(1-\cos q_n)^2. \label{disp}
\end{equation}

\begin{figure}[t]
\epsfig{file=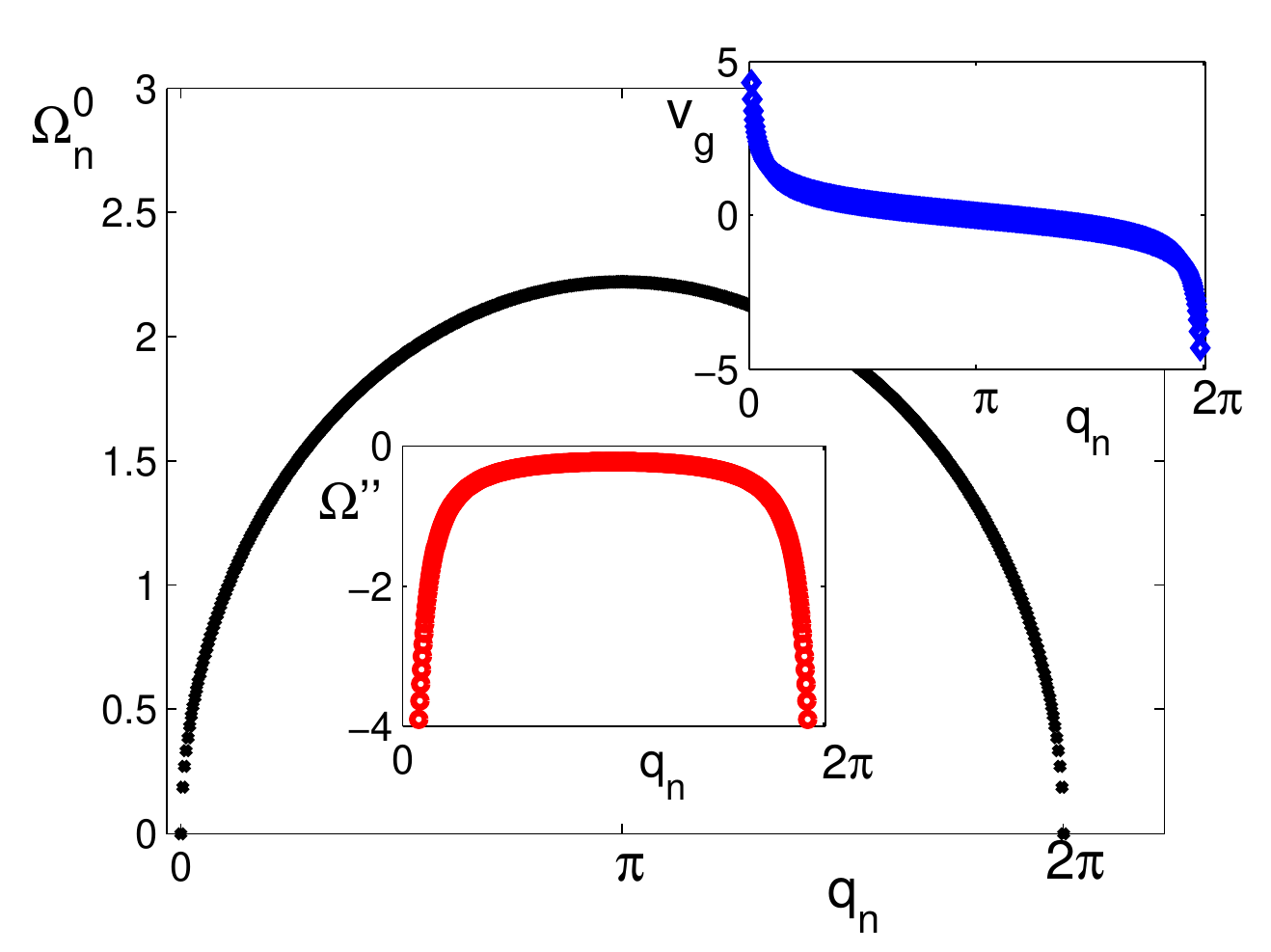,width=1\linewidth} \caption{Main plot: dispersion
relation~(ref{disp}) in the linear limit $A\rightarrow 0$ for
discretized values of the wavenumber $q_n=2\pi n/N$, $N=512$. Top
inset: the first discrete derivative of the dispersion $v_g$ (group
velocity). Lower inset: the second discrete derivative $\Omega
\prime\prime$.} \label{lattice1}
\end{figure}

This dispersion relation in the linear limit $A\rightarrow 0$ is
shown in the main plot of Fig.~\ref{lattice1}. The insets show its first
derivative $v_g$ (group velocity) and second derivative $\Omega^{\prime\prime}$.
Derivatives are performed by discrete
differences for a sufficiently large finite value of $N$ with step size
$h=2\pi/N$. In all calculations below we will the use power-law exponent
$\alpha=2$. Both the group velocity $v_g$ and $\Omega^{\prime\prime}$ diverge
when $N \to \infty$ in the zero wavenumber limit $q_n \to 0$.

Let us concentrate first our attention on the solitons that appear at small wavelength.
As usual~\cite{boardman}, we represent the solution as an expansion in normal
modes
\begin{equation}
u_\ell =\frac{1}{2}\left[\sum\limits_{n=1}^N C_n
e^{i(q_n\ell-\Omega_n t)}+c.c.\right].
\end{equation}
Focusing on the carrier wavenumber $q^0$ of the wave packet and the
associated frequency $\Omega^0$ in the limit $A\rightarrow 0$, i.e.
$\Omega^0\equiv\Omega(q^0,A=0)$, and defining
$\Omega_n=\Omega^0+\epsilon^2\delta\Omega_n$,
$q_n=q^0+\epsilon\delta q_n$, we get an expression of the form of
Eq.~(\ref{plane})
\begin{equation}
\label{def}
u_\ell= \epsilon\left[A(\zeta,\tau)e^{i(q^0\ell-\Omega^0
t)}+c.c.\right].
\end{equation}
where the envelope function $A(\zeta,\tau)$, is a slowly
varying function in space $\zeta = \epsilon\ell$ and time $\tau = \epsilon^2t$.
Its expression is
\begin{equation}
\label{A}
A(\zeta,\tau)=\sum\limits_n C_n e^{i(\delta q_n \zeta-\delta\Omega_n
\tau)}.
\end{equation}

Taking the time derivative of formula~(ref{A}) we get
\begin{equation} \label{dA}
\frac{\partial}{\partial \tau}A(\zeta,\tau)=\sum\limits_n C_n
e^{i(\delta q_n\zeta-\delta\Omega_n
\tau)}\left[-i\delta\Omega_n(q_n)\right].
\end{equation}
Expanding now $\epsilon^2\delta\Omega_n(q_n)=\Omega_n(q_n)-\Omega^0(q^0)$ around
the value $q_n^0$ and $A=0$ we get approximately
\begin{equation}
\delta\Omega_n(q_n)=\sum_{s=1}^\infty\frac{(\epsilon\delta q_n)^s}{
\epsilon^2s!}\frac{\Delta_h^{(s)}\left[{\Omega^0(q^0)}\right]}{h^s}+|A|^2\frac{
\partial\Omega_n({ q^0})}{\partial(|A|^2)}\biggr|_{A=0} \label{tay}
\end{equation}
where $\Delta_h^{(s)}$ is the difference operator of order $s$ with
step size $h=2\pi/N$ in the limit $A=0$. We will restrict ourselves
to difference operators of first and second order:
$\Delta_h^{(1)}{\left[\Omega^0\right]\equiv\Omega^0(q^0+h)-\Omega^0(q^0)}$
and
$\Delta_h^{(2)}{\left[\Omega^0\right]\equiv\Omega^0(q^0+h)+\Omega^0(q^0-h)-2\Omega^0(q^0)}$.
Then, substituting~(ref{tay}) into~(ref{dA}) and taking into account
the definition~(\ref{A}), we get the following nonlinear equation
for the amplitude $A(\zeta,\tau)$

\begin{figure}[t]
\epsfig{file=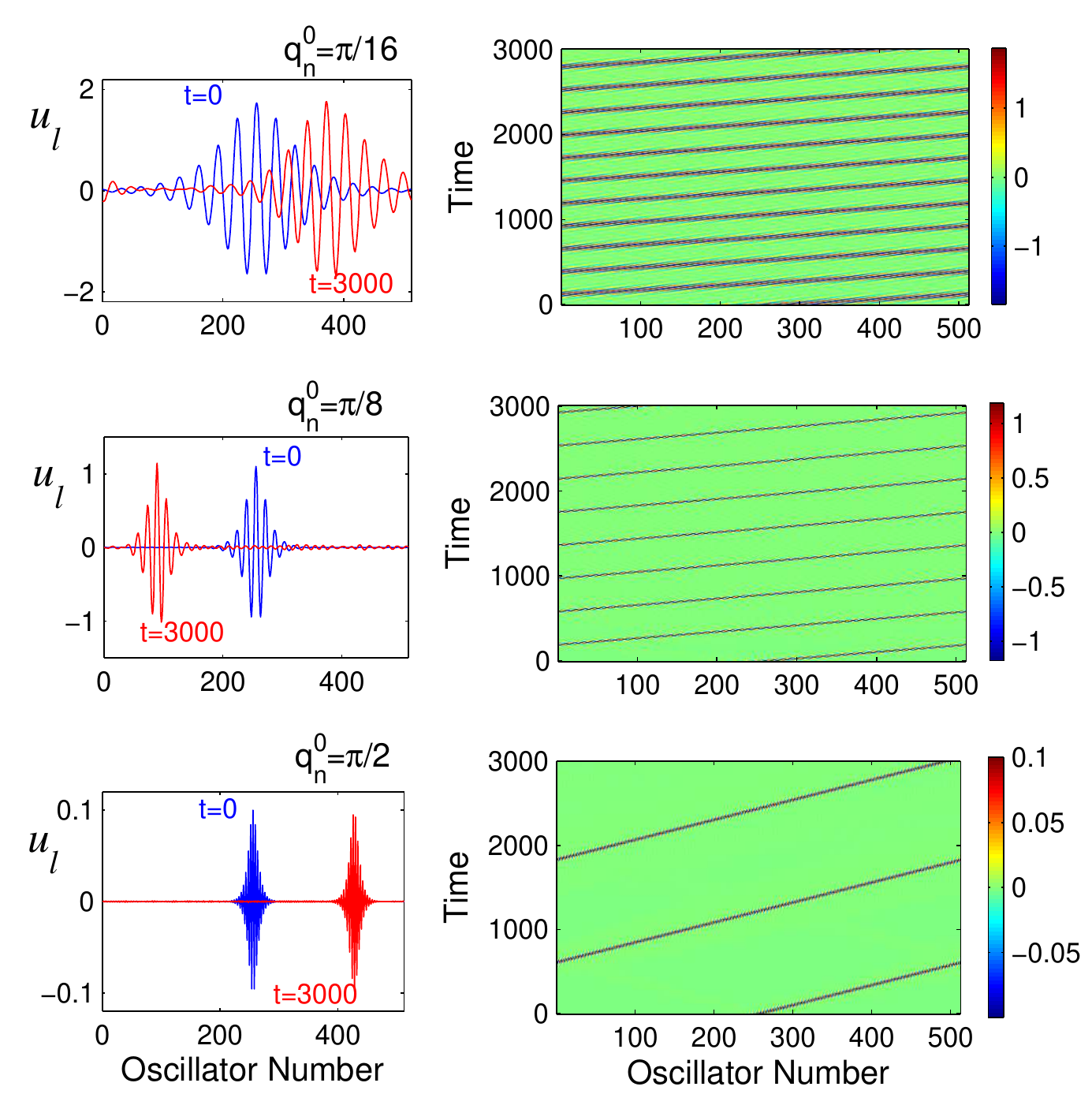,width=1\linewidth} \caption{Left panel: Displacement field $u_l$ as a
function of site number $l$ at times t=0 (blue) and t=3000 (red) for different
carrier wavenumbers: $q^0=\pi/16,\pi/8$ and $\pi/2$ (from top to bottom). Right
panel: Space-time evolution of the displacement field $u_k$ represented in color code.}
\label{lattice}
\end{figure}

\begin{equation}
\label{non}
i\left(\frac{\partial A}{\partial \tau}+\frac{v_g}{
\epsilon}\frac{\partial A}{\partial {
\zeta}}\right)+\frac{\Omega''}{2}\frac{\partial^2 A}{\partial {
\zeta}^2}-R|A|^2A=0,
\end{equation}

where the definitions below have been introduced
\begin{eqnarray}
\label{param}
&v_g=\Delta_h^{(1)}\left[{\Omega^0}\right]/h, \qquad
\Omega''=\Delta_h^{(2)}\left[{\Omega^0}\right]/h^2, \\
&R=\partial\Omega_n({ q^0})/\partial(|A|^2)\Bigr|_{A=0}={ 3(1-\cos
q^0)^2/2\Omega^0}. \nonumber
\end{eqnarray}

Now, switching to a comoving reference frame with a rescaled time $t$
and space $\xi$ where $\xi={ \epsilon}(\ell-v_gt)$, we get the Non Linear
Schr\"odinger (NLS) equation

\begin{equation}
\label{33}
i\frac{\partial A}{\partial t}+\frac{\Omega''}{2} \frac{\partial^2
A}{\partial\xi^2}-R|A|^2A=0
\end{equation}
This equation has the well known one-soliton solution for $A$, which
can be inserted in expression~(ref{def}) to get

\begin{equation}
\label{soliton2} u_\ell(t) = \frac{f\cos\left[\left({ \Omega^0}+
\frac{3(1-\cos { q^0})^4}{4{ \Omega^0}}f^2\right)t - { q^0}
\ell\right]}{\cosh\left[f(1-\cos { q^0})^2
\sqrt{\frac{3}{2\Omega_n^0 |\Omega^{\prime\prime}|}}(\ell-v_g
t)\right]},
\end{equation}
where $f$ stands for the soliton amplitude.

It is noteworthy to mention that the semi-discrete
approach~\cite{short,nika,jam} based on the continuous reductive
perturbation theory~\cite{taniuti} fails in the derivation of the
soliton profile~(ref{soliton2}). In particular, a derivation similar
to Ref.~\cite{short} leads to the same NLS equation~(\ref{33}), but
with a different dispersion coefficient containing continuous
derivatives $\Omega''=\partial^2 { \Omega^0/\partial q^2}$ instead
of difference operators as given in Eq.~(\ref{param}). The
appearance of continuous derivatives causes a divergences of both
the group velocity $v_g$ and $\Omega^{\prime\prime}$. Indeed,
considering continuous derivatives
\begin{equation}
\label{disp1}
\frac{\partial^2 { \Omega^0}}{\partial q^2}=
\frac{1}{{\Omega^0}}\sum_{m=1}^N \frac{\cos(q
m)}{m^{\alpha-2}}-\frac{1}{[{\Omega^0}]^3}\left(\sum_{m=1}^N
\frac{\sin(q m)}{m^{\alpha-1}}\right),
\end{equation}
the first term on the right-hand side is divergent for $\alpha=2$.
Specifically, it oscillates as a function of both $q$ and $N$. In
summary, the approach in Refs.~\cite{short,nika,jam,taniuti} does
not lead to the correct expression for the soliton parameters. The
correct approach relies on a discrete wave-packet
dynamics~\cite{boardman} and on the use of discrete difference
operators, as done in this Letter.

\section{Results}
We have performed numerical simulations of the set of
equations~(ref{first}) with periodic boundary conditions, using
solution~(\ref{soliton2}) as an initial condition ($t=0$), i.e. we
consider $u_\ell(t=0)$ as initial displacements and $\dot
u_\ell(t=0)$ as initial velocities. In Fig~\ref{lattice} we display
the time-evolution of this initial condition for three different
wavenumbers, approaching $q^0=0$ from bottom to top. As predicted by
linear theory the group velocity $v_g$ increases when $q^0$
decreases. At the same time, the width of the soliton grows as well
and thus the wave amplitude must increase in order to keep the
soliton within the lattice length limits.

\begin{figure}[t]
\epsfig{file=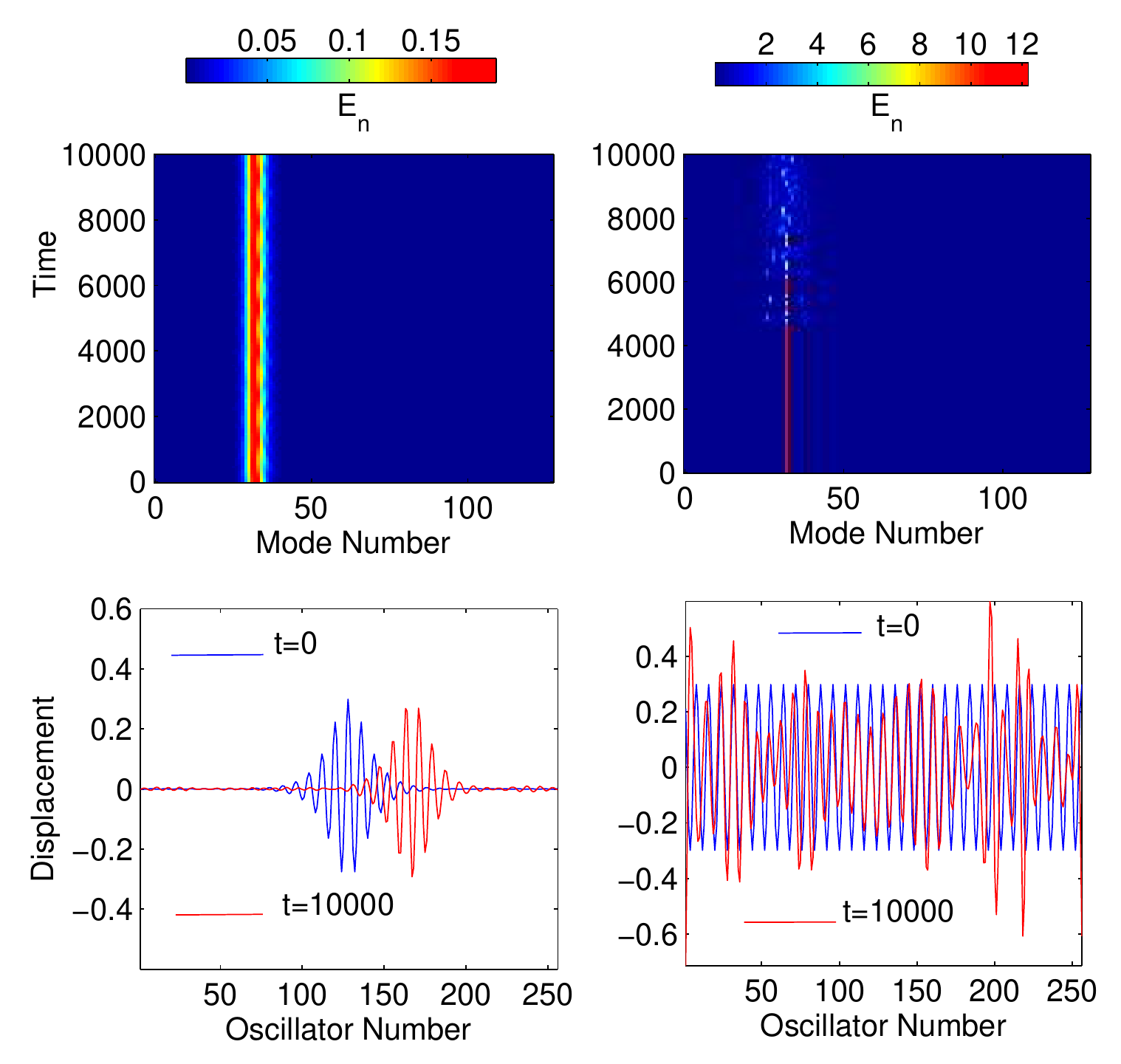,width=1\linewidth} \caption{Left panels: Time
evolution of solitonic solution~(ref{soliton2}) in mode space (top)
and in real space (bottom). Right panel: Time evolution of the
single mode in mode space (op) and in real space (bottom). For
better visualization only the first 125 modes are displayed, all
other modes carry almost zero energy.} \label{fast}
\end{figure}

Traveling envelope solitons are robust against perturbations and we
do not observe their destruction on a long time scale, while single
carrier mode excitation with the same amplitude is modulationally
unstable. This instability is presented in Fig.\ref{fast}. The
soliton shape (left panel) remains unchanged up to the time
$t=10^4$, while the single mode excitation (right panel) collapses
on a much shorter time interval.

One has to mention that the effective nonlinearity parameter in our
model is $f\sqrt{R/|\Omega^{\prime\prime}|}$ and we can therefore
increase the wave amplitude $f$ in the small carrier wavenumber
limit without violating the weak nonlinearity restriction. Beyond
this weak nonlinearity limit, traveling envelope solitons become
unstable or they are trapped by the lattice and transform into
standing breather solutions~\cite{flach}.

The existence of other weakly nonlinear localized solutions, like kink-solitons, is limited
by the fast increase of the group velocity in the low wavenumber limit. In fact,
low wavenumber excitations, which should in principle
generate kink-solitons, are characterized by drastically different
velocities and they cannot form localized wave packets obeying
the Korteweg-de Vries (KdV) or the modified KdV equations~\cite{zabusky,kos1,poggi,short}.
These problems appear as well in the case of the analytic description of envelope solitons.

\begin{figure}[t] \center
\epsfig{file=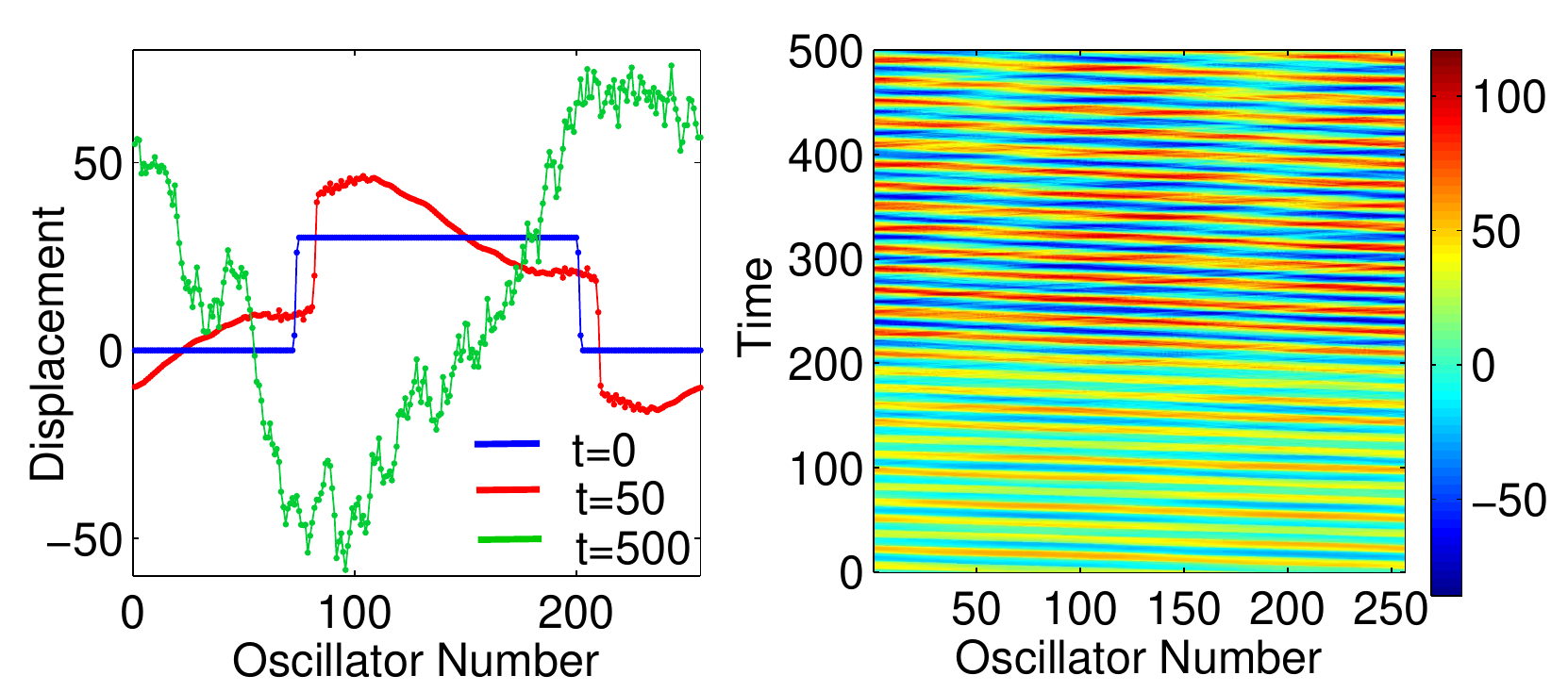,width=1\linewidth} \caption{Left panel:
Displacement pattern with initial kink-antikink solution~(ref{kink})
with amplitude $A=20$ at different time: $t=0$, $t=50$ and $t=500$.
Right panel: Color code representation of the evolution of the
displacement pattern in space-time of the same kink-antikink pair.
As clearly seen, the kink-antikink pair survives up to time $t=200$
after which the system goes to a chaotic regime.} \label{fast2}
\end{figure}

However, at strong nonlinearity, kink-soliton solutions with "magic"
wavenumber $2\pi/3$ exist. Extended waves with that particular
wavenumber are exact solutions of model~(ref{first}). It has been
proposed that such exact solutions can acquire a compact support and
maintain their validity as approximate solutions~\cite{kos2,ram3}.
These truncated wave solutions can be written in terms of relative
displacements as follows
\begin{equation}
u_{\ell+1}-u_\ell=\pm\frac{A}{2}\left[1+\cos\left(\frac{2\pi}{3}\ell-\omega
t\right)\right] \label{kink}
\end{equation}
for $\left|2\pi \ell/3-\omega t\right|<\pi$ and
$u_{\ell+1}-u_\ell=0$ for $\left|2\pi \ell/3-\omega t\right|>\pi$,
where
\begin{equation}
\omega\equiv\Omega(q=2\pi/3,A)=\sqrt{(\Omega_{N/3}^0)^2+(45/16)A^2}
\label{w}
\end{equation}
and the supersonic kink is characterized by the group velocity
$v=3\omega/2\pi$. We have considered here these approximate
analytical kink-antikink solutions~(ref{kink}) as initial
conditions. As far as we use periodic boundary conditions, it is
impossible to consider single kink motion: we have thus monitored
the dynamics of kink-antikink pairs (see Fig.~\ref{fast2}). As it
appears from numerical simulations, although the dynamics follows
approximately the solution~(ref{kink}), the kinks are much less
robust against the collisions with perturbative excitations than it
happens in the case of envelope solitons. At the beginning the kink
shape remains unchanged, but the kink-antikink motion creates
perturbations in the lattice and those inhomogeneities finally cause
the destruction of the kink solution.

\section{Conclusions}
Concluding, we have analytically found moving soliton solutions in a
long-range version of the FPU model~(ref{first}). Those are weakly
nonlinear envelope solitons and strongly nonlinear kink-soliton
solutions. Envelope solitons show stable propagation along the
lattice at variance with kink-solitons which collapse on a short
time scale. Numerical simulations confirm the validity of the
analytic approximate solutions.

\section{Acknowledgements}

The work is supported in part by contract LORIS (ANR-10-CEXC-010-01). R. Kh. acknowledges financial support from
Georgian SRNSF (grant No FR/25/6-100/14) and travel grants from
Georgian SRNSF and CNR, Italy (grant No 04/24) and CNRS, France
(grant No 04/01).

\end{document}